\documentclass[aps,pra,showpacs,preprintnumbers,amsmath,amssymb]{revtex4}
\usepackage{graphicx}
\usepackage{bm}
\usepackage{dcolumn}

\def\nat#1#2#3{Nature {\bf #1}, #2 (#3)}

\def\sc#1#2#3{Science {\bf #1}, #2 (#3)}
\def\prv#1#2#3{Phys. Rev. {\bf #1}, #2 (#3)}
\def\rmp#1#2#3{Rev. Mod. Phys. {\bf #1}, #2 (#3)}
\def\prl#1#2#3{Phys. Rev. Lett. {\bf #1}, #2 (#3)}

\def\prb#1#2#3{Phys. Rev. B {\bf #1}, #2 (#3)}
\def\prd#1#2#3{Phys. Rev. D {\bf #1}, #2 (#3)}

\def\pop#1#2#3{Phys. Plasmas {\bf #1}, #2 (#3)}

\def\noi{\noindent}
\def\bc{\begin{center}}
\def\ec{\end{center}}
\newcommand{\bea}{\begin{equation}}
\newcommand{\eea}{\end{equation}\noi}
\newcommand{\ber}{\begin{eqnarray}}
\newcommand{\eer}{\end{eqnarray}\noi}
\begin{document}
\title{Fermi-Dirac Statistics}
\author{Shyamal Biswas}\email{sbsp [at] uohyd.ac.in}
\affiliation{School of Physics, University of Hyderabad, CR Rao Road, Gachibowli, Hyderabad-500046, India}

\date{\today}

\begin{abstract}
Here we have discussed on Fermi-Dirac statistics, in particular, on its brief historical progress, derivation, consequences, applications, etc. Importance of Fermi-Dirac statistics has been discussed even in connection with the current progresses in science. This article is aimed mainly for undergraduate and graduate students.

\end{abstract}
\pacs{01.65.+g, 03.75.Ss, 05.30.-d, 01.55.+b}
\maketitle
\section{Introduction and Brief Historical Progress}
Fermi-Dirac statistics describes energy distribution of a non (or weakly) interacting gas of identical particles (now called fermions, eg. neutrino, electron, quark, proton, neutron, $^{6}$Li, $^{40}$K, \textit{etc}) that obey the Pauli exclusion principle \cite{pauli}. It is named after Enrico Fermi who derived it in 1926 and Paul Dirac who derived it independently a little later in the same year \cite{fermi,dirac}. It was a very important time for the development of quantum mechanics as well as of modern physics particularly for the revolutionary transition from the old quantum theory to the new quantum theory. While the old quantum mechanics mostly was commanded by Bohr-Sommerfeld quantization \cite{bohr,sommerfeld}, new quantum mechanics is commanded by Schrodinger's wave mechanics \cite{schrodinger} and Heisenberg-Born-Jordan matrix mechanics \cite{heisenberg}. This time spin ($1/2$) of an electron although was proposed, yet not well understood \cite{pauli2}. Nonetheless, the Pauli exclusion principle worked- it clearly explained the structure of the periodic table \cite{mendeleev}, the atomic spectra (fine structure \cite{fine-structure}, anomalous Zeeman effect \cite{zeeman} and Paschen-Back effect \cite{paschen-back}), \textit{etc}. 

While for a single-particle system, an operator (eg. Hamiltonian) is quantized using the (first) commutation relation between position and momentum, a successful attempt of quantizing the (energy) distribution of particles, for a many-particle system, was made by Fermi using Pauli exclusion principle as another (second) quantization rule for identical particles not even well knowing about the new quantum mechanics. Dirac, of course, was well aware of new quantum mechanics when he obtained a similar distribution formula. Fermi and Dirac essentially established a connection between (classical) Maxwell-Boltzmann distribution \cite{maxwell-boltzmann} and (quantum) Pauli exclusion principle \cite{pauli}. They perhaps were motivated to do such a brilliant work from the qualitative prediction of Nernst on the `degeneracy' of gases at low temperature \cite{nernst}. `Degeneracy' of the non (or weakly) interacting (Fermi) gas, of course, is a natural outcome of their brilliant connection. 

It should be mentioned in this regard, that, a similar brilliant successful attempt (towards  `degeneracy' of a gas at low temperature) for another type of gas of identical particles (now called bosons eg. photon, gluon, $\pi^0$, $^1$H, $^4$He, $^7$Li, $^{14}$N, $^{16}$O, $^{23}$Na, H$_{2}$O, $^{52}$Cr, W$^\pm$ bosons, $^{87}$Rb, Z-boson, Higgs boson, \textit{etc}) was also made by Bose and Einstein even earlier than Fermi did \cite{bose,einstein}. Spin of particles, other than electron, was of course, not known at that time. Spin of photon was observed to be $1$ in 1931 by Raman and Bhagavantam \cite{raman}. Connection between spin of identical particles and (Fermi-Dirac or Bose-Einstein) statistics was known from the spin statistics theorem proposed in around 1940 by Fierz and Pauli \cite{spin-statistics}. Categorization of atoms to either bosons or fermions started around this time \cite{london,atoms}. Later on, quantum field theory came to us as a complete! description of a many particle system interpreting interparticle 
interaction among fermions as exchange of bosons eg. exchange of photon leads to Coulomb interaction between two electrons \cite{feynman}, exchange of Z-boson leads to weak interaction between electron and neutrino \cite{weinberg}, exchange of pi-meson ($\pi^0$) or gluons leads to strong interaction between proton and neutron \cite{strong-interaction1} or more strongly among quarks \cite{strong-interaction2}, \textit{etc}. That, atoms of integral spin obey Bose-Einstein statistics, was confirmedly verified much later in 1995 \cite{bec}, and that, atoms of half integral spin obey Fermi-Dirac statistics, was verified even later in 1999 \cite{fermi-degeneracy}. However, historical development of Fermi-Dirac statistics as well as Pauli exclusion principle was mainly associated with that of atomic spectral lines. Let us now discuss on some previous history in this regard.   

Existence of spectral lines was known to us since the beginning of the nineteenth century when Wollaston and Fraunhofer first observed the dark absorption lines in the spectrum of the Sun \cite{wollaston-fraunhofer}. It took almost eighty years, the time necessary to improve the quality of technical instruments, to build-up an empirical law for the hydrogen spectral lines as observed in stars. In 1885 (even before the discovery of electron by Thomson \cite{thomson} and that of proton by Rutherford \cite{rutherford-proton}) Balmer showed that the wavelengths of the visible spectral lines of hydrogen atom could be represented by a simple mathematical formula $\lambda=h\frac{n_{2}^2}{n_{2}^2-n_{1}^2}$, where $\lambda$ is wavelength, $h=3645.6$ \AA, $n_1=2$, and $n_2=3, 4, 5, 6,...$ respectively for the first, second, third, and so forth elements of the series \cite{balmer}. A few years later after this, Rydberg generalized Balmer's empirical formula as $\frac{1}{\lambda_n}=R_{\infty}\big(\frac{1}{(n_1+\mu_1)^2}-\frac{1}{(n_2+\mu_2)^2}\big)$, where $\lambda_n$ is the wavelength of the n-th member of the series, $R_{\infty}\approx1.1\times10^7$m$^{-1}$ is the Rydberg constant, and $\mu_1$ denotes spectral terms (e.g. $\mu_1\rightarrow^3{\bf{P}}_1$, $\mu_2\rightarrow^3{\bf{S}}_1$) \cite{rydberg}. Putting $\mu_1=\mu_2=0$, $n_1=2$ and $n_2=3,4,5,...$ in Rydberg's phenomenological formula, one can easily get Balmer's empirical formula. Rydberg probably was the first person to distinguish a \textit{sharp} series from a \textit{principal} series or \textit{diffuse} series. Other type of series (\textit{fundamental}) was discovered later in around 1907 \footnote{For very short review, see W.B. Jensen, \textit{The origin of s,p,d,f orbitals}, J. Chem. Educ. \textbf{84}, 757 (2007)}. However, Rydberg's phenomenological formula did not offer a theoretical understanding of atomic spectra. The first step towards such an understanding taken was Bohr's atomic theory of hydrogen proposed in 1913 \cite{bohr}. Bohr's semiclassical theory for stability 
of hydrogen atom was compatible with Planck's quantum hypothesis \cite{planck}, Einstein's theory on photoelectric effect \cite{photoelectric-effect}, and with Rutherford's model for atoms \cite{rutherford-model}. Consequently, he obtained Rydberg's formula in an specialized form $\frac{1}{\lambda_n}=R_{\infty}\big(\frac{1}{n_{1}^2}-\frac{1}{n_{2}^2}\big)$, where $n_1, n_2=1,2,3,4...$, $n_1<n_2$, and the Rydberg constant this time was obtained in terms of fundamental constants as $R_\infty=\frac{m_ee^4}{64\pi^3\epsilon_0^2\hbar^3c}$. With the touch of quantization, Rutherford model became familiar with the name: shell model or simply the Bohr model. Bohr's theory was later generalized (even for other potentials) by Sommerfeld resulting a name called Bohr-Sommerfeld quantization \cite{sommerfeld}. Following up on Sommerfeld, at the beginning of the 1920s Bohr revised the shell model for atoms (further introducing an atomic core model), and developed a building-up schema (aufbau principle) for the elements of 
Mendeleev's periodic table \cite{bohr-pt,mendeleev}. Although Pauli exclusion principle was implicitly there in the building-up schema, Bohr could not uniquely predict the (electronic) ground states of the atoms in the periodic table \footnote{Exclusion principle was also proposed less precisely by E.C. Stoner [\textit{The distribution of electrons among atomic levels}, Philos. Mag. \textbf{48}, 719 (1924)] even before Pauli.}. Bohr-Sommerfeld semiclassical theory could not also explain fine structure \cite{fine-structure}, anomalous Zeeman effect \cite{zeeman}, Paschen-Back effect \cite{paschen-back}, \textit{etc}. These problems challenged the old (Bohr-Sommerfeld) quantum mechanical theory. As a remadey, the (new) quantum mechanics (Heisenberg's uncertainty principle, Heisenberg-Born-Jordan matrix mechanics \cite{heisenberg}, Schrodinger equation \cite{schrodinger}, and Dirac's mathematical formulation of quantum mechanics \cite{dirac-qm}), spin \cite{pauli2}, Pauli exclusion principle \cite{pauli}, Hund's rules \cite{hund}, \textit{etc} came into the picture as new quantum revolution in around 1926 \footnote{Later on Lamb's 
spectroscopic experiment on hydrogen atom challenged another revolution to come in around 1948 with a (new) quantum field theory \cite{lamb,feynman}.}. Fermi-Dirac statistics \cite{fermi,dirac}, like Bose-Einstein statistics \cite{bose,einstein}, was another important addition to the new quantum revolution, in particular, for the non (or weakly) interacting gas of identical particles which obey both the Pauli exclusion principle \cite{pauli} for all temperatures ($T$) and the Maxwell-Boltzmann distribution \cite{maxwell-boltzmann} for $T\rightarrow\infty$. 

\section{Derivation and Consequences}
Although Fermi-Dirac statistics originally was derived for a microcanonical ensemble of ideal Fermi gas for harmonically trapped case \cite{fermi} and relativistic! case \cite{dirac}, yet the original treatments were generalized later on for canonical and grandcanonical ensembles \footnote{Ensemble picture was introduced by J.W. Gibbs, \textit{Elementary principles in statistical mechanics}, Charles Scribner's Sons, New York \& Edward Arnold, London (1902).}. For convenience, let us outline the natural treatment for a grandcanonical ensemble. 

Let us consider an ideal gas (macroscopic system) of identical particles which obey Pauli exclusion principle. Let is be connected with a thermal \& particle reservoir which is characterized by temperature $T$ and chemical potential $\mu$. In equilibrium, grand potential ($\Phi_G$) of the system attains its minimum value, and consequently, temperature of the system becomes $T$ and chemical potential of the system becomes $\mu$. Since our system is composed of identical particles, individual Hamiltonian for all the particles are the same. If we further consider that all the particles have the same spin component (say spin $\uparrow$) then all of them are indistinguishable. If half of the particles carries $\uparrow$ component and the other half carries $\downarrow$ component, then first half of the particles are distinguished from the rest. For simplicity, let us consider that all the particles have the same spin component. Let eigenvalues of the single particle Hamiltonian be given by $\epsilon_0, \epsilon_1,
 \epsilon_2, \epsilon_3,...,\epsilon_j,...$, and number of particles to the complete set of orthonormal states ($\{|j>\}$) be $n_0, n_1, n_2, n_3,..., n_j,...$. For our system, each possible set of the numbers ($\{n_j\}$) is a microstate. In statistical mechanics, in a grand canonical ensemble of identical thermodynamic systems, the equilibrium grand potential (grand free energy) for a single thermodynamic system of identical particles is postulated as
\begin{equation}
\Phi_G=-k_BT\ln Z_G,
\end{equation}
where
\begin{equation}
Z_G=\sum_{\{n_j\}}\text{e}^{-\sum_{j}(\epsilon_j-\mu)n_j/k_BT}
\end{equation}
is the grand partition function for the system. For grandcanonical ensemble, there is no restriction of total number of particles. So, we can do the sum over $\{n_j\}$ in Eqn.(2) as a product of sum over each individual $n_j$. On the other hand, Pauli exclusion principle (no two indistinguishable particles can occupy a single state) restricts the value of each individual $n_j$ as either $0$ or $1$. Thus we recast Eqn.(2) as
\begin{equation}
\Phi_G=-k_BT\sum_{j}\ln\big(1+\text{e}^{-(\epsilon_j-\mu)/k_BT}\big).
\end{equation}
Statistical mechanics also postulates the probability of a microstate ($\{n_i\}$) of the thermodynamic system in the grandcanonical ensemble as
\begin{equation}
P_{\{n_i\}}=\frac{\text{e}^{-\sum_{i}(\epsilon_i-\mu)n_i/k_BT}}{Z_G}.
\end{equation}
Here-from it is very easy to get average number of particles at a single-particle state $|j>$ for the thermodynamic system as
\begin{equation}
\bar{n}_j=\sum_{\{n_i\}}n_jP_{\{n_i\}}=\frac{\partial\Phi_G}{\partial\epsilon_j}\bigg|_{T,\mu}=\frac{1}{\text{e}^{(\epsilon_j-\mu)/k_BT}+1}.
\end{equation}
This is the celebrated Fermi-Dirac statistics. One of the postulates of statistical mechanics is as follows: all the ensemble averages, calculated according to the scheme of statistical mechanics, must match in the thermodynamic limit with their respective thermodynamic quantities. This postulate allows us to write, that, in the thermodynamic limit, above formula would be the same as that originally obtained for microcanonical ensemble \cite{fermi,dirac}.    

For a free Fermi gas, $\epsilon_j\rightarrow\epsilon_{\textbf{p}}=\text{p}^2/2m$, where $\textbf{p}$ is the momentum and $m$ is the mass of each particle. Let volume of this system be $V$, (average) number of particles be $N$, number density be $\bar{n}=N/V$, and pressure be $p$, then according to Gibbs-Duhem thermodynamic relation \cite{duhem}, we can easily write $\Phi_G=-pV$. Total (average) number of particles and equation of state can now be given by $N=\sum_{\textbf{p}}\bar{n}_{\textbf{p}}=\frac{V}{\lambda_T^3}f_{3/2}(z)$ and $p=-\Phi_G/V=\bar{n}k_BT\frac{f_{5/2}(z)}{f_{3/2}(z)}$, where $z=\text{e}^{\mu/k_BT}$ is the fugacity and $f_{j}(z)=z-z^2/2^j+z^3/3^j-...$ is a Fermi function. Average energy per particle can be obtained as $U=\frac{1}{N}\sum_{\textbf{p}}\epsilon_{\textbf{p}}\bar{n}_{\textbf{p}}=\frac{3k_BT}{2}\frac{f_{5/2}(z)}{f_{3/2}(z)}$. While $z\rightarrow0$ denotes the classical limit ($T\rightarrow\infty$), $z\rightarrow\infty$ denotes the `quantum' limit ($T\rightarrow0$). For 
asymptotically power-logarithmic behavior of the Fermi functions, it can be shown that, as $T\rightarrow0$, the pressure of the free Fermi gas, for a fixed number density, approaches a nonzero constant ($p\rightarrow\frac{2}{5}\epsilon_F\bar{n}$) unlike that ($p\rightarrow0$) obtained from the classical Maxwell-Boltzmann distribution formula \cite{sommerfeld-metal} \footnote{$\epsilon_F$ is the Fermi energy, and for ideal homogeneous Fermi gas of a single spin component, it is $\epsilon_F=\frac{\hbar^2}{2m}(6\pi^2\bar{n})^{2/3}$.}. This nonzero pressure is called the Pauli pressure. It is a consequence of Pauli exclusion principle. Bulk modulus of a metal mostly is a result of Pauli pressure. Pauli pressure also resists the gravitational collapse of a (lighter!) white dwarf or neutron star. In the low temperature regime ($k_BT/\epsilon_F\ll1$), specific heat per particle ($c_v=\frac{\partial U}{\partial T}|_{V,N}$) can be obtained using Fermi-Dirac statistics as $c_v=k_B\frac{\pi^2}{2}\big(\frac{k_BT}{\epsilon_F}\big)^1+{\it{O}}\big(\frac{k_BT}
{\epsilon_F}\big)^3$ \cite{sommerfeld-metal} which is far deviated from the classical result ($c_v^{\text{(cl)}}=3k_B/2$), and is compatible with the Nernst's heat theorem (hypothesis) \cite{third-law} and matches favourably with the experimental data \cite{cv-measurement} \footnote{The first mathematical calculation towards the justification of Nernst's hypothesis (which later was extended as the third law of thermodynamics) was presented by A. Einstein, \textit{Planck's theory of radiation and the theory of the specific heat}, Ann. Physik \textbf{22}, 180 (1907) [in German].}. This was a great success of Fermi-Dirac statistics over the Maxwell-Boltzmann statistics in the low temperature regime. This form of specific heat, unlike its classical form, also leads to the physically acceptable value of Lorenz number for metal as $L=\frac{K}{\sigma T}=\frac{\pi^2}{3}\big(\frac{k_B}{e}\big)^2\approx2.45\times10^{-8}$ watt-ohm/deg$^2$, where $K$ is the thermal conductivity and $\sigma$ is the electrical conductivity \cite{ssp}. 

In presence of magnetic field ($\textbf{B}=B\hat{k}$) energy levels of an electron can be written as $\epsilon_{p_z,j}=p_z^2/2m+2\mu_BB(j+1/2)\mp\mu_BB$ where the first term represents free motion along $z$ direction, the second term represent circular motion along $x-y$ plane, third term represents breaking of spin degeneracy, and $j=0,1,2,3,...$ represents Landau level. For this set of energy levels, Fermi-Dirac statistics leads to the grand potential as \cite{biswas}
\begin{eqnarray}
\Phi_{G}&=&-k_BT\sum_{p_z,j}\ln(1+\text{e}^{-(\epsilon_{p_z,j}-\mu)/k_BT})\nonumber\\&&=-Nk_BT\frac{f_{5/2}(z)}{f_{3/2}(z)}+\text{Im}\bigg[\frac{4Nk_BTb}{f_{\frac{3}{2}}(z)\sqrt{\pi}}\sum_{k=1}^\infty\frac{1}{2\pi k}\int_0^\infty\big[\text{e}^{\frac{k\pi^2}{b}}\text{B}(-\text{e}^{y-\nu},\nonumber\\&&\frac{ik\pi}{b},0)+i\pi\text{csch}(\frac{k\pi^2}{b})\big]\text{e}^{i\pi k(\nu-y)/b}y^{-1/2}\text{d}y\bigg],
\end{eqnarray}
where $b=\mu_BB/k_BT$, $\nu=\mu/k_BT$, and $\text{B}(-\text{e}^{y-\nu},\frac{ik\pi}{b},0)$ is an incomplete beta function. Now from $\chi=-\frac{\mu_0}{V}\frac{\partial^2\Phi_G}{\partial B^2}$, one can easily get magnetic susceptibility ($\chi$) of the ideal gas of electrons. 

For a spherically symmetric case of 2:1 mixture of ideal homogeneous gas of relativistic electrons and interacting gas of relatively static helium ions in a star, relativistic energy of an electron is given by $\epsilon_{\textbf{p}}=\sqrt{p^2c^2+m_e^2c^4}$, which, in the ultrarelativistic and low temperature regime ($pc\gg m_ec^2$, $T/T_F\rightarrow0$ \footnote{$T_F=\epsilon_F/k_B$ is the Fermi temperature.}), leads to the total effective kinetic energy of the mixture as $E_k=\sum_{\textbf{p}}\epsilon_{\textbf{p}}\bar{n}_{\textbf{p}}=\frac{3V}{2\Lambda^3}\big[\frac{cp_F^4}{4}+\frac{m_e^2c^3p_F^2}{4}+...\big]$, where $\Lambda=(3\pi^2)^{1/3}\hbar$ and $p_F=\Lambda(\bar{n}/2)^{1/3}$ is the Fermi momentum. Potential energy of this system can effectively be given by $E_p=-\frac{3}{5}\frac{GM^2}{R}$, where $M=2N_em_p$ is the total effective mass, $N_e$ is the total number of electrons, $(4/3)\pi R^3=V$ is the volume. The equilibrium radius of the mixture can now be set as the minimum of the total energy ($E=E_k+E_p$), which leads to the critical 
mass of the mixture as $M_C\sim\frac{(\hbar c/G)^{3/2}}{m_p^2}$ \cite{huang}. This way one can explore stability of the white dwarf stars. Fermi-Dirac statistics has so many other/related consequences as mentioned in the next section.

\begin{figure}
\includegraphics{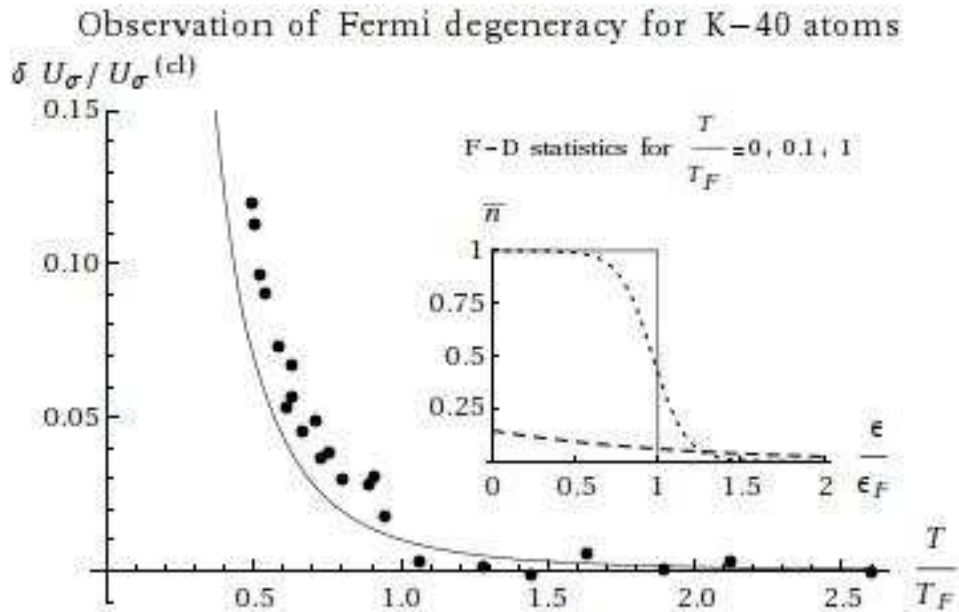}
\caption{Solid line represents ideal gas prediction for excess energy ($\delta U_\sigma=U_\sigma-U_\sigma^{(\text{cl})}$) of harmonically trapped Fermi gas, $U_\sigma=\frac{1}{N_\sigma}\sum_{i}\epsilon_i\bar{n}_i=3k_BT\frac{f_4(z)}{f_3(z)}$ is the average energy per particle of spin $\sigma=\pm1/2$, $N_\sigma$ is the total number of particles of spin $\sigma$, and $U_\sigma^{(\text{cl})}=3k_BT$ is the average classical energy per particle \cite{fermi-degeneracy}. Circles represent experimental data adapted from \cite{fermi-degeneracy}. Solid, dotted and dashed lines in inset respectively represent Fermi-Dirac statistics (Eqn.(16)) for $T/T_F=0$, $0.1$ and $1$.}
\end{figure}

\section{Application}
Since any particle is either boson or fermion, and fermions take part in forming structure of matter, Pauli exclusion principle as well as Fermi-Dirac statistics must have huge applicability. That, the conduction electrons of the metals can be thought of as an ideal gas of electrons, is a reason for Fermi-Dirac statistics to be applied for understanding of the thermal conductivity, electrical conductivity, superconductivity \cite{superconductivity}, magnetic susceptibility, classical \& quantum Hall effects \cite{hall,quantum-hall}, \textit{etc} \cite{ssp}. 

Fowler probably was the first person to correctly applied the Fermi-Dirac statistics for a distinct practical problem. In late 1926, he proposed that the relationship among the density, energy and temperature of a white dwarf star could be explained by viewing them as an ideal nonrelativistic gas of electrons and nuclei which obey Fermi-Dirac statistics \cite{fowler}. He also predicted about the ultimate fate of a white dwarf star considering the Fermi gas model. The Fermi gas model was then used by Frenkel, Anderson and Stoner in 1928-30 to calculate the relationship among the mass, radius, and density of the white dwarf stars, assuming them to be essentially homogeneous spheres of electron gas \cite{frenkel-anderson-stoner}. Soon after this, Chandrasekhar obtained a value of critical mass ($M_C\approx1.4M_{\text{O}}$ \footnote{Here $M_{\text{O}}\approx2\times10^{30}$ kg is the solar mass.}) for the stability of the white dwarf star applying Fermi-Dirac statistics for an ideal inhomogeneous gas of relativistic 
electrons \cite{chandrasekhar} \footnote{Chandrasekhar probably was not aware of the previous works of Frenkel, Anderson, Stoner and others when he made the draft of the paper in around 1930.}. Around this time, applying Fermi-Dirac statistics (possibly not knowing the previous works of Chandrasekhar and others), Landau obtained the value of critical mass not only for the white dwarf stars but also for the `neutron' stars (even before the discovery of neuron by Chadwick \cite{neutron-discovery}) \cite{landau-neutron}. Most possibly, it was the first application of Fermi-Dirac statistics to a system of composite fermions (neurons). 

Thomas-Fermi model which now-a-days is commonly used for approximation technique (even for Bose systems) was proposed as an application of Fermi-Dirac statistics in 1927 \cite{thomas-fermi}. Fermi gas nuclear model was also proposed by Majorana and Weizsacker as a similar application for the description binding energy of the nucleons in a nucleus in around 1934 \cite{majorana-weizsacker}.   

In 1927, applying Fermi-Dirac statistics for the conduction electrons, Pauli explained weak temperature dependence of the paramagnetic susceptibility ($\chi=\frac{3}{2}\frac{\mu_B^2\bar{n}}{\epsilon_F}-\it{O}(B^2)$) of metals exposed in a weak magnetic field ($\textbf{B}=B\hat{k}$) \cite{pauli-paramagnetism}. This explanation was compatible (to a certain extent) with the existing experimental observations \cite{magnetism-experiment}, and confirmed that, (conduction) electrons obey Fermi-Dirac statistics, and each of the electrons itself is a tiny (spin) magnet with two fold degeneracy (in absence of magnetic field). Such an application modernized the theory of metals from Drude-Lorentz classical model \cite{drude-lorentz} to Sommerfeld-Bloch semiclassical model \cite{sommerfeld-metal,bloch} in around 1929. Soon after this, a diamagnetic and oscillatory effect on the magnetization for conduction electrons was added by Landau \cite{landau-dia} to the paramagnetic contribution obtained by Pauli to explain 
experimental observations on magnetization of metals particularly for strong magnetic field and to open a theory for quantum Hall effect \cite{quantum-hall-theory}. In 1931, electronic band structure came as a very important addition  to explain metallic, semiconductivity and insulator behavior of a crystalline solid body \cite{kronig-penney}. Around this time, for the consequence of application of Fermi-Dirac statistics, existence of `hole' was proposed by Heisenberg \cite{heisenberg2} and that of positron was proposed by Dirac \cite{dirac2}. Like 
electrons and positrons, `holes' also obey Fermi-Dirac statistics. This way Fermi-Dirac statistics found its area of applicability greatly to the theory of metals and semiconductors \cite{ssp}.

Spin statistics theorem was necessary to be proposed by Fierz and Pauli in around 1940 to connect spin of particles to either Bose-Einstein statistics or Fermi-Dirac statistics \cite{spin-statistics}. This theorem extended and generalized Dirac's canonical quantization technique for photons to open quantum field theories for system of identical particles \cite{dirac3}. Consequently, quantum hydrodynamic theory (for Bose liquid) \cite{landau-qht}, quantum electrodynamics \cite{feynman}, Landau's Fermi liquid theory \cite{landau-flt}, BCS theory on superconductivity \cite{bcs}, Abrikosov flux lattice \cite{type-2}, Anderson localization \cite{anderson-localization}, standard model \cite{weinberg}, asymptotically free gauge theory \cite{strong-interaction2}, \textit{etc} came to us as direct or indirect applications of Fermi-Dirac statistics to revolutionize our theoretical understanding of physics. From 1950s Fermi-Dirac statistics has majorly been applied to the development of quantum field 
theories for many particle systems in particular towards condensed matter and particle physics. This statistics has so many other or related applicability in statistical mechanics, low temperature physics, nuclear physics, semiconductor physics, low dimensional physics, plasma physics, astrophysics, supersymmetric theory \cite{susy}, grand unified theory \cite{gut}, string theory \cite{string}, carbon nano tube physics \cite{cnt}, mesoscopic physics \cite{spin-hall}, ultra-cold atom physics \cite{fermi-degeneracy,ucap}, graphene physics \cite{graphene}, topological insulator physics \cite{t-i}, \textit{etc}.

\section*{Acknowledgment}
We are thankful to S. Chaturvedi of University of Hyderabad for his valuable comments and suggestions. In-fact this article has some overlap with \textit{Fermi-Dirac statistics: derivation and consequences}, S. Chaturvedi and S. Biswas, Resonance \textbf{19}, 45 (2014). Useful discussions with J.K. Bhattacharjee of HRI are gratefully acknowledged. We are also thankful to Saugata Bhattacharyya of Vidyasagar College for supplying some useful references from his prestigious collection. This work was sponsored in part by the Department of Science and Technology [DST], Govt. of India under the INSPIRE Faculty Award Scheme {[No. IFA-13 PH-70]}.


\begin{thebibliography}{99}
\bibitem{atoms}Atoms or even protons/neutron are composed of fundamental particles (quarks, electrons). They are either composite bosons or composite fermions according to the addition of spins of the constituent fundamental particles.
\end{thebibliography}
\end{document}